\documentclass{iaus}
\usepackage{graphicx}

\title[JD 11.~~Composition of Giant Planets] 
{Composition of Massive Giant Planets}

\author[Ravit Helled$^1$, Peter Bodenheimer$^2$, and Jack J. Lissauer$^3$]   
{Ravit Helled$^1$, Peter Bodenheimer$^2$, 
 \and Jack J. Lissauer$^3$}

\affiliation{$^1$Department of Earth and Space Sciences, \\University of California, Los Angeles, CA 90095-1567, USA,
\\
$^2$University of California, Santa Cruz, CA 95064, USA\\
$^3$ NASA-Ames Research Center, Moffett Field, CA 94035, USA \\email: {\tt rhelled@ucla.edu or r.helled@gmail.com}}

\pubyear{2010}
\volume{276}  
\pagerange{119--126}
\jname{The Astrophysics of Planetary Systems: Formation, Structure, and Dynamical Evolution}
\editors{}
\begin{document}

\maketitle

\begin{abstract}
The two current models for giant planet formation are core accretion
 and disk instability. We discuss the core masses and overall planetary enrichment in
heavy elements 
 predicted by the two formation models, and show that both models could lead
 to a large range of final compositions.  For example, both can form giant planets with nearly stellar compositions. However, 
low-mass giant planets, enriched in heavy elements compared to their host
 stars, are more easily explained by the core accretion model. The final
 structure of the planets, i.e., the distribution of heavy elements, is not
 firmly constrained in either formation model. 
\end{abstract}

\firstsection 
\section{Introduction}

The topic of giant planet formation has been studied for decades. 
The discoveries of giant planets outside our solar system have given us the
opportunity to test existing formation and interior models of gas giant
 planets. Transiting extrasolar giant planets provide important information
 about their bulk compositions, and a large range of compositions for
these objects has been deduced (Guillot, 2008).
Observations of extrasolar gas planets have raised many challenges for giant
 planet formation theories, including the determination of planetary
 formation location, formation timescale, and planetary composition and 
structure. 
The interplay between theories and observations is expected 
to provide a clearer and more complete picture of giant planet formation. \par

The standard model for giant planet formation, {\it `core accretion'}, is
 based on the hypothesis that the formation of a giant planet begins with 
planetesimal coagulation and heavy-element core formation, followed by
 accretion of a gaseous envelope 
 (Pollack et al., 1996; Lissauer et al., 2009).
A second model for giant planet formation is {\it `gravitational
 (disk) instability'} in which gas giant planets form as a result of 
gravitational fragmentation in the disk surrounding the young star 
(Boss, 1997; Durisen et al., 2007). \par 

There are substantial differences between the two formation models, 
including formation timescale, the most favorable formation location,  and 
the ideal disk properties for planetary formation. The different nature of the
 two formation mechanisms naturally leads to some expected differences in 
planetary composition. However, the final composition of a planet formed by
 each of these models depends upon the local disk
 properties  in the region of planetary birth. As a result, no clear-cut 
criterion  regarding  planetary composition can be used to discriminate
 between these two formation mechanisms. Below we discuss the ranges of 
 planetary compositions predicted by the two models, and their possible
 differences.

\section{Core Accretion}
In the core accretion model, first, dust grains accrete to form solid planetesimals that merge to form
 a solid core surrounded by a thin gaseous atmosphere. Solids, which 
 typically consist of rocks, ices and organics, are accreted on a time  scale
of 1 to a few Myr, while the gas accretion rate initially falls well below
the solid accretion rate.
The solid accretion rate decreases significantly once the planetesimals
 in the planet's feeding zone are depleted, while the gas accretion rate 
increases steadily. Eventually, the gas accretion rate exceeds the accretion
 rate of solid planetesimals, and gas continues to be accreted at a nearly 
constant rate. Once the core mass  and the mass of the gaseous envelope 
 become about equal, a runaway gas accretion builds up the mass of the envelope
rapidly while leaving the core at a  nearly constant value. Gas accretion then
 stops either by dissipation of nebular gas or by gap opening (Lissauer et al., 2009). After
 that point the planet is practically isolated from the disk, and it contracts
 quasi-statically  and cools on a time scale of $10^9$ years (see Lissauer
 \& Stevenson, 2007 and references therein).      

\subsection{Core Mass}
The core masses of giant planets formed by core accretion can vary from a 
few Earth masses (M$_{\oplus}$) up to even tens of M$_{\oplus}$. 
The growth of the core continues as long as planetesimals are present in the
 planet's feeding zone. The planet's feeding zone represents the region around
 the planet from which it can capture planetesimals, and its extent is 
normally taken to be about 4 Hill sphere radii inward and outward from the 
planet's orbit.
The Hill sphere radius of the protoplanet is 
$R_H=a\left(\frac{M_p}{3M_{\star}}\right)^{1/3}$ where $a$ is the planet's 
semimajor axis, $M_p$ is the planet's 
mass, and $M_{\ast}$ is the mass of the star. 
Once the feeding zone is depleted, the planetary object is nearly isolated and
 planetesimal accretion drops significantly. The mass at which isolation occurs
 is given by 
\begin{equation}
M_{iso}\approx  \frac{64}{\sqrt{3}} \pi^{3/2} M_\star^{-1/2} \sigma^{3/2} a^3
\label{eq:miso}
\end{equation}
where  $\sigma$ is the solid surface density (Lissauer, 1993).  Around a 1 M$_\odot$  star at 5.2
AU with $\sigma = 10$ g cm$^{-2}$, this mass is about 11.5 M$_\oplus$.

The final core mass $M_c\approx \sqrt{2} M_{iso}$ (Pollack et al. 1996) because
the feeding zone for planetesimals expands during the slow gas accretion
phase. Equation (\ref{eq:miso}) relates the core mass and disk properties provided that neither the planet nor the planetesimals migrate substantially. 
 First, the isolation mass increases
with radial distance, so planets in wider orbits are predicted to have more
 massive cores. Second, the isolation mass is proportional to $\sigma^{3/2}$ (Lissauer, 1987). 
 The more solids available to the growing planet, 
the more massive the core can become. It is therefore clear that the core mass
 would increase with increasing $\sigma$, and that for a
 given disk mass (or metallicity), steeper density profiles would result in
 massive cores at smaller radial distances (see Helled \& Schubert, 2009 for
 details). Finally, $M_{iso}$ is inversely proportional to the square root of
stellar mass. Note that Movshovitz et al. (2010) find that core accretion can produce planets with cores of only a few earth masses.

\subsection{Planetary Composition}
The core accretion model does not predict a specific
 composition for the forming planet. The total heavy-element enrichment of the
 planet depends on (1) the planetary core mass, which, as just discussed,
depends on disk properties, (2) on the amount of dust entrained within 
 the accreted gaseous envelope, and (3) planetesimals (or even other planets)
 accreted subsequent to rapid gas accretion. \par

The total mass fraction of heavy elements in the planet can by given by 
$Z$ = $(M_{c}+M_{Z_{env}})/M_{p}$, where $M_{Z_{env}}$ is the mass of heavy
 elements in the gaseous atmosphere.
This simple relation demonstrates the dependence of the bulk composition on
 both the core mass and the abundance of the accreted gas. 
 The relation implies that the final composition of a massive planet is strongly dependent on the accreted gas composition. 
 \par

For relatively massive planets, the core contains only a very small fraction
 of the total mass. The predicted planetary $Z$ depends on the
 accreted gas composition which could be that of the star, or
 dust-free or dust-enriched relative to the star.
 This point is illustrated in Fig. 1a, for
different values of $M_c$, assuming that the composition of the gaseous
envelope is solar. However, planets which are depleted
 in heavy
 elements compared to their host star could form as well. The gaseous envelope
 that is accreted could be depleted with solids 
 leading to a sub-stellar $Z$. 
 Giant planets that are enriched with heavy elements compared to the star
 could form when cores are massive, when the accreted gas is enriched
 with solids, and when planetesimals are accreted at later stage of the planetary growth. It is therefore clear that the core accretion model could lead
 to the formation of planets with a wide range of final masses and compositions.

\begin{figure}[h!]
    \centering
        \includegraphics[width=5.895in]{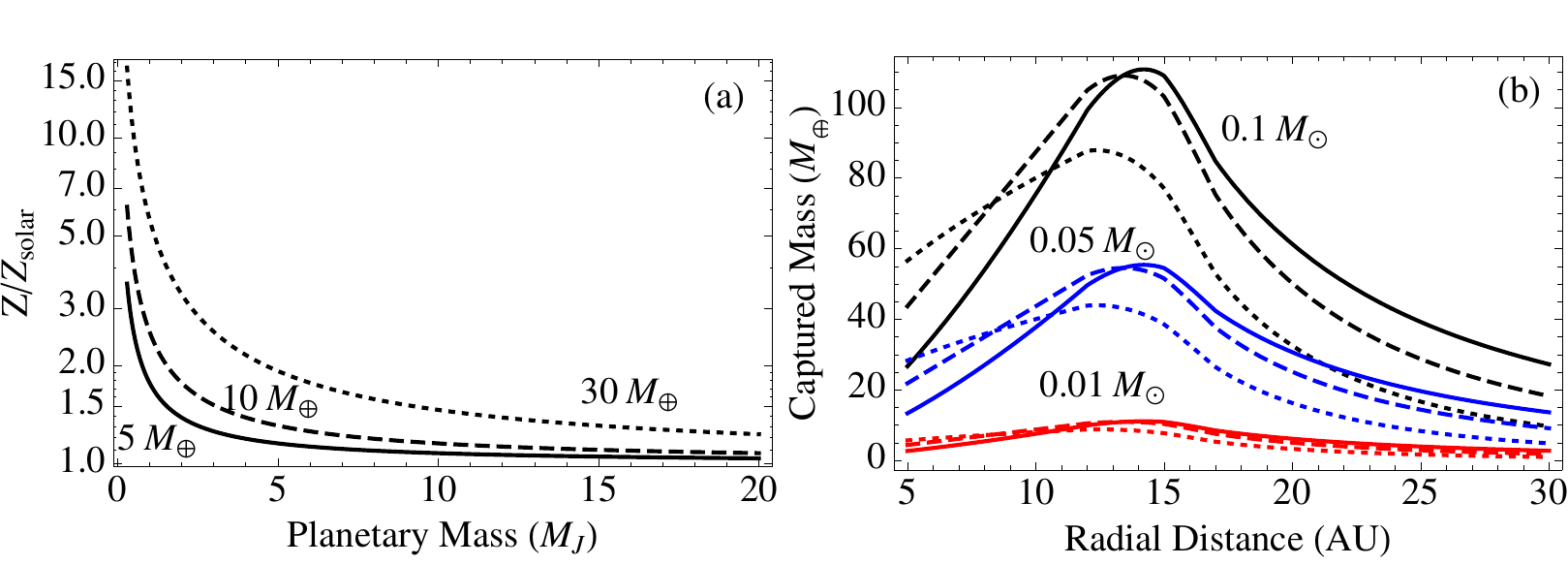}
    \caption[err]{({\bf a}):  metallicity of the planet divided by solar metallicity as a function of planetary mass
 for three different core masses: 5, 10, and 30 M$_{\oplus}$ (dotted, dashed,
 and solid curves, respectively) assuming  a solar composition for the gas.
({\bf b}): Captured solid mass by a planet formed by gravitational
instability, as a function of the distance of the planet from the star.
The black, blue
and red curves are for disk masses of 0.1, 0.05, and 0.01 M$_{\odot}$,
 respectively. The solid, dashed, and dotted curves refer to density
 distributions proportional to $a^{-1/2}$, $a^{-1}$ and $a^{-3/2}$, 
respectively (Helled \& Schubert, 2009).}
\end{figure}

\section{Gravitational Instability}
In the disk-instability model, planets are formed by fragmentation of the gas
 disk due to gravitational instabilities. Once a local instability occurs,
 a gravitationally bound sub-condensation region can be created. The fragment
 contracts, and eventually evolves to become a giant planet. 
Numerical investigations suggest that planets in wide  orbits, such as those recently 
observed by direct imaging (Marois et al., 2008), could form by disk
 instability.  

 The evolution of planets formed by gravitational
 instability is often followed using standard stellar (planetary)
 evolution equations, under the approximation of 
spherical symmetry.  The model typically takes the objects to be
 gravitationally bound,  isolated, homogenous and static, with stellar
 composition (Bodenheimer et al., 1980; Helled et al., 2006).\par

The first evolutionary stage is the `pre-collapse'
 stage in which the extended ($\approx 0.5$ AU) and cold (internal
temperatures $< 2000$ K) protoplanet contracts quasi-statically.
 The duration of this stage is not well constrained  but as a first approximation it can be taken to be inversely
 proportional to the square of the initial mass of the body (Helled \& Bodenheimer,
 2010b), so more
 massive protoplanets
 evolve faster. The pre-collapse stage ends when molecular hydrogen starts to
 dissociate at the center of the body and a hydrodynamic collapse ensues 
(Bodenheimer et al., 1980).  
\par

\subsection{Planetary Composition}
The initial composition of giant planets formed by gravitational instability is
 similar to the young disk's composition and is therefore stellar. However, 
 the planet may subsequently be enriched in heavy elements by planetesimal capture (Helled et al., 2006; Helled \& Schubert, 2008). \par

The amount of solid mass available for capture depends on the solid surface density
  $\sigma$ at
 the planetary location. As in the core accretion model, $\sigma$ changes with
 the disk mass and its radial density profile, and with stellar metallicity. 
The total available mass of solids in the planet's feeding zone can be given by
\begin{equation}
M_{av}= 16 \pi a^2  \left(\frac{M_p}{3M_{\star}}\right)^{1/3}\sigma. 
\label{eq:mav}
\end{equation}
Eq. (\ref{eq:mav}) shows that the available mass for capture on one hand
 increases with radial distance due to its dependence on $a^2$, but also
 decreases with radial distance due to its dependence on $\sigma$, which
 decreases  as a function of $a$. It is therefore clear that the mass available 
 for capture strongly depends on the planetary semimajor axis $a$, and the disk
 properties (Helled \& Schubert, 2008).  Also, the available mass depends
weakly on $M_p$. The available mass for accretion
 provides an upper limit to the enrichment for given conditions assuming that the planetesimals and/or the protoplanet do not migrate substantially. 
 However, the
 actual enrichment depends on the ability of the planet to capture these
 solids during its pre-collapse contraction. \par

The planetesimal accretion rate is given by (Safronov, 1969)
\begin{equation}
\frac{dm}{dt}=\pi R^2_{cap} \sigma \Omega F_g
\label{eq:saf}
\end{equation}
where $R_{cap}$ is the protoplanet's capture radius, $\Omega $ is the
 protoplanet's orbital frequency, and $F_g$ is the gravitational enhancement
factor.
Eq. (\ref{eq:saf}) suggests that the accretion rate is significantly smaller
at large radial distances due to its dependence on both $\sigma$ and $\Omega$.
As a result, giant planets at wide orbits will have insufficient time for accretion, 
which leads to negligible enrichment of solids (Helled \& Bodenheimer, 2010a).
 The protoplanet's capture radius depends on the planetary size and density,
 and is therefore larger with increasing planetary mass. Finally, the accreted 
mass depends on the available time for accretion. As discussed in Helled
 \& Schubert (2008), enrichment is efficient as long as the protoplanet is
 extended and fills most of the area of its feeding zone,
 so planetesimals can be slowed down by gas drag, and be absorbed by the 
protoplanet. Therefore the longer the pre-collapse stage is, the longer the 
time available for accretion; thus low mass protoplanets have
 more time to accrete solids, although the mass available for accretion is limited because of  its dependence on
the total planetary mass. \par

The planetary enrichment can therefore change significantly from one system to
 another. Helled et al. (2006) have shown that  a Jupiter-mass (M$_J$) clump
 formed at 5.2 AU could have accreted more than 40 M$_{\oplus}$ of heavy
 elements. 
Helled \& Schubert (2009) have shown that a 1 M$_J$ protoplanet formed between 
5 and 30 AU could accrete 1-110 M$_{\oplus}$ of heavy elements, depending
on disk properties, and concluded that in the disk instability model the final
composition of a giant planet is strongly determined by its formation
environment. Fig. 1b shows the heavy element enrichment found by Helled 
\& Schubert (2009). The figure presents the captured mass in the first 10$^5$
years of the planetary evolution for all the cases considered. 

Helled \& Bodenheimer (2010a) have found that protoplanets with masses between
 3 and 7 M$_J$ in wide orbits (24 to 68 AU) can accrete between tens and
 practically zero M$_{\oplus}$, with the negligible mass corresponding to the
 larger radial distances. 
In the disk instability model, it is predicted that the planetary enrichment
 with heavy elements strongly decreases with radial distance, and that giant
 planets in wide orbits would have nearly stellar composition. 
 
\subsection{Core Mass}
In the disk instability model the newly formed protoplanets have no cores.
 However, cores could form by settling of solids towards the planetary center 
(DeCampli \& Cameron, 1979). 
Helled et al. (2008) presented a detailed analysis of coagulation and
 sedimentation of silicate grains in an evolving 1 M$_J$ protoplanet including
 the presence of convection. It was found that during the initial contraction
 of the protoplanet (pre-collapse stage), which lasts several times
 $10^5$ years,
 silicate grains can sediment to form a core both for convective and
 non-convective envelopes, although the sedimentation time is substantially 
longer if the envelope is convective. Grains made of ices and organics were 
found to dissolve in the planetary envelope, not contributing to core formation. \par 
 
Helled \& Schubert (2008) have investigated the topic of core formation for
 protoplanets with masses between 1 Saturn mass and 10 M$_J$. 
It was found that grain settling occurs in low-mass protoplanets which are
 cold enough and have a contraction time-scale long enough for the grains to
 grow and sediment to the center. The grain sedimentation process was found to
 be favorable for low-mass bodies, due to lower internal temperatures, lower 
convective velocities and longer contraction time-scales. In convective
 regions, the grains are carried by the convective eddies until they 
grow to
 sizes of 10 cm or larger (the exact value depends on the convection velocity 
which changes with depth, time and planetary mass). Then they are massive enough to
 decouple from the gas, sediment to the center and form a core. Protoplanets
 with masses of $\geq$ 5 M$_{J}$ were found to be too hot to allow core
 formation. 
In this mass range the grains evaporate in the
planetary envelope, enriching it with refractory material (see Helled \&
 Schubert, 2008 for details). \par

As a result, in the disk instability model, $M_c$ is {\it not} simply
 proportional to the mass of the protoplanet. Low-mass clumps have the lowest
 convective velocities and longest evolution--properties that support
 core formation. The final core mass depends on the available solid mass within
 the planet's envelope. If a substantial amount of solids can be accreted 
before internal temperature are high enough to evaporate the grains, cores of
 several M$_{\oplus}$ could be formed (Helled et al., 2006; Helled \& Schubert,
2008). 

\section{Internal Structure}
Neither model for giant planet formation can provide tight constraints on
 the planetary internal structure. 
In core accretion, one prediction would be that the core consists of rocks,
 organics, and ices, while in the disk instability model the cores would be
 composed of refractory material. Thus, if large pieces of ice and organic
 material enter the planetary envelope, they could settle to the
 core as well (Helled et al., 2006; Helled \& Schubert, 2008). In addition,
 core accretion simulations show that once the core mass reaches 
$\sim$3 M$_{\oplus}$, volatiles in small solid bodies dissolve in the envelope and do not reach
 the core, while for simplification, the model typically assumes that the
 accreted planetesimals fall to the core. Also, it was found that a substantial
amount of ices could stay in the envelope (Iaroslavitz \& Podolak, 2007). \par

Both of these formation models, although advanced in many ways, do not follow
 throughout the evolution the fate of the high-$Z$ material. For example,
 accreted planetesimals could either fall to the center or evaporate in the
 envelope and mix with the gas, grains can slowly settle to the center, and
 high-$Z$ material from the core could be mixed back up by core erosion. These
 processes (and more) suggest that the final internal structure of a giant
 planet is not well constrained, and cannot be directly related to the 
formation mechanism. \par

In the context of extrasolar giant planets, one must therefore think in terms
 of global or bulk composition, instead of planetary structure or atmospheric
 enrichment. Even for Jupiter and Saturn, for which the gravitational 
moments are
 measured, the internal structures are not well determined, and are often
 dependent on model  assumptions, equations of state, etc. 
At present, the available information on the compositions of extrasolar giant 
planets is insufficient for discriminating between the two models for giant
 planet formation. Providing predictions of the internal structures, and
 atmospheric properties of giant planets is still very challenging for
 formation models.  

\section{References} 
\noindent Bodenheimer, P., Grossman, A. S., Decampli, W. M., Marcy, G. \& Pollack,
 J. B., 1980. Icarus 41, 293.\\
Boss, A. P., 1997. Science 276, 1836.\\
Decampli, W. M., \&  Cameron, A. G. W., 1979. Icarus, 38, 367.\\
Durisen, R. H., Boss, A. P., Mayer, L.;,Nelson, A. F., Quinn, T. \& Rice, W. K. M., 2007.
In Protostars and Planets V, B. Reipurth, D. Jewitt, and K. Keil (eds.),
 Univ. of Arizona Press, Tucson,  607.\\
Guillot, T., 2008. Nobel Symposium 135,
 Stockholm, Physica Scripta, 130, 014023, arXiv:0712.2500.\\
Helled, R.  \& Bodenheimer, P., 2010a. Icarus, 207, 503.\\
Helled, R. \& Bodenheimer, P., 2010b. Icarus, in press.\\
Helled, R., Podolak, M. \& Kovetz, A., 2006. Icarus, 185, 64.\\
Helled, R., Podolak, M. \& Kovetz, A., 2008. Icarus, 195, 863. \\
Helled, R.  \& Schubert, G., 2008. Icarus, 198, 156.\\
Helled, R. \& Schubert, G., 2009. ApJ, 697, 1256.\\
Iaroslavitz, E.,  \& Podolak, M., 2007. Icarus, 187, 600.\\
Lissauer, J. J., 1987. Icarus, 69, 249.\\
Lissauer, J. J., 1993. Annu. Rev. Astron. Astrophys. 
31, 129.\\
Lissauer, J. J.,  \& Stevenson, D. J., 2007. In 
Protostars and Planets V,   B. Reipurth, D. Jewitt, and K. Keil (eds.),
 University of Arizona Press, Tucson,  591.\\
 Lissauer, J. J., Hubickyj, O., D'Angelo, G. \& Bodenheimer, P., 2009. Icarus, 199, 338.\\ 
Marois, C., Macintosh, B., Barman, T., Zuckerman, et al., 2008.
Science
322, 1348. \\
Movshovitz, N., Bodenheimer, P., Podolak, M. \& Lissauer, J. J., 2010. Icarus, 209, 616.\\
Pollack, J. B., Hubickyj, O., Bodenheimer, P., Lissauer, J. J., Podolak, M.,
 \& Greenzweig, Y., 1996. Icarus, 124, 62.\\
Safronov, V. S., 1969. Evolution of the Protoplanetary Cloud and the Formation of the Earth and Planets. Nauka, Moscow. \\


\end{document}